\documentclass{aa}

\usepackage{graphicx}
\usepackage{txfonts}
\usepackage{dsfont}
\usepackage{url}
\usepackage[breaklinks,colorlinks,citecolor=blue]{hyperref}

\usepackage{natbib}
\bibpunct{(}{)}{;}{a}{}{,}

\newcommand\be{\begin{equation}}
\newcommand\ee{\end{equation}}
\newcommand\bea{\begin{eqnarray}}
\newcommand\eea{\end{eqnarray}}
\def\ba#1\ea{\begin{align}#1\end{align}}

\newcommand\dd{\mathrm{d}}

\newcommand\mr{\mathrm}
\newcommand\lbra{\left\langle}
\newcommand\rbra{\right\rangle}
\newcommand\Cov{\mr{Cov}}

\newcommand\Clgal{\ensuremath{C_\ell^\mathrm{gal}}}
\newcommand\nbargal{\ensuremath{\overline{n}_\mathrm{gal}}}

\newcommand\lmax{\ell_\mathrm{max}}
\newcommand\kmax{k_\mathrm{max}}
\newcommand{\threeJz}[3]{\begin{pmatrix} #1 & #2 & #3 \\ 0 & 0 & 0 \end{pmatrix}}

\newcommand{\Euclid}{\textit{Euclid}}
\newcommand{\SKA}{\textit{SKA}}
\newcommand{\uuidSKA}{e1ab3737-721f-493f-9d03-2e27115ecdcd}

\begin{document}

\title{Cosmology in the non-linear regime : the small scale miracle}
\titlerunning{Small Scale Miracle}

\author{Fabien Lacasa \thanks{fabien.lacasa@u-psud.fr} \inst{\ref{inst1},\ref{inst2}} }
\institute{
Institut d’Astrophysique Spatiale (IAS), Bâtiment 121, F-91405 Orsay, Université Paris-Sud 11 and CNRS, UMR 8617, France\label{inst1}
\and
D\'{e}partement de Physique Th\'{e}orique and Center for Astroparticle Physics, Universit\'{e} de Gen\`{e}ve, 24 quai Ernest Ansermet, CH-1211 Geneva, Switzerland\label{inst2}
}

\date{\today}

\abstract
{
Interest rises to exploit the full shape information of the galaxy power spectrum, as well as pushing analyses to smaller non-linear scales. Here I use the halo model to quantify the information content in the tomographic angular power spectrum of galaxies $C_\ell^\mr{gal}(i_z)$, for future high resolution surveys : \Euclid{} and \SKA{}2. I study how this information varies as a function of the scale cut applied, either with angular cut $\lmax$ or physical cut $\kmax$.  For this, I use analytical covariances with the most complete census of non-Gaussian terms, which proves critical. I find that the Fisher information on most cosmological and astrophysical parameters follows a striking behaviour. Beyond the perturbative regime we first get decreasing returns : the information keeps rising but the slope slows down until reaching a saturation. The location of this plateau is a bit beyond the reach of current modeling methods : $k\sim 2$ Mpc$^{-1}$ and slightly depends on the parameter and redshift bin considered. I explain the origin of this plateau, which is due to non-linear effects both on the power spectrum, and more importantly on non-Gaussian covariance terms. 
Then, pushing further we see the information rising again in the highly non-linear regime, with a steep slope. This is the small scale miracle, for which I give interpretation and discuss the properties. Hints are shown that this information should be disentanglable from the astrophysical content, and could improve Dark Energy constraints. Finally, more hints are shown that high order statistics may yield significant improvements over the power spectrum in this regime, with the improvements increasing with $\kmax$.
Data and notebooks reproducing all plots and results will be made available at \url{https://github.com/fabienlacasa/SmallScaleMiracle}
}

\maketitle


\section{Introduction}\label{Sect:intro}

Future surveys of the large scale structure such as \Euclid{} \citep{Euclid-redbook}, \textit{LSST} \citep{Abell:2009aa} and \SKA{} \citep{Maartens:2015mra} will allow high resolution mapping of the distribution of galaxies in the Universe. Exploiting the most out of these data sets would require to (i) use the full shape of the statistical measurements, in contrast for instance with targeted BAO extraction, and (ii) push the analyses to the smallest accessible scales.

Full shape information of the galaxy power spectrum can indeed extract information from faint features \cite[e.g.][]{Tansella2018} but also from the general slope (to constrain $n_S$), and from the amplitude (to constrain $\sigma_8$ and the growth of structure) if used in conjunction with weak lensing or higher order correlation \citep{Hoffmann2015a}. This has been shown to encode more constraining power than usual BAO and RSD analyses \citep{Loureiro2019,Troster2019}.

Pushing to small scales is challenging for future surveys because they are entering the non-linear regime of structure formation where the physics of the dark matter halos become relevant. There is however a wealth of evidence that the halo properties do encode cosmological information to constrain Dark Energy and Gravity \citep{Balmes2014,Lopes2018,Lopes2019,Contigiani2019,Ryu2019}. This motivates rising interest to use non-linear scales for cosmological constraints \citep[e.g.]{Lange2019}.

For the matter field, various methods have been developed to predict P(k) to smaller scales with the required 1\% precision, for instance the \Euclid{} emulator can reach $k\sim$ 1 h/Mpc \citep{EuclidEmulator2019} for $\Lambda$CDM, and fitting functions are proposed to push to $k\sim 10/\mr{Mpc}$ \citep{Hannestad2019}. This 1\% target can also be reached for models beyond $\Lambda$CDM --including Dark Energy, Modified Gravity and neutrinos-- through rescaling methods based on the halo model \citep{Cataneo2019a,Giblin2019,Cataneo2019b}.

For galaxy clustering, the situation is complexified by the galaxy formation physics. Halo model-based approaches are however showing their strength to extract cosmology out of galaxy statistics either with machine learning methods \citep{Ntampaka2019}, or with analytical Halo Occupation Distribution \citep{Kobayashi2019}. Emulators have also been developed, and for instance the \texttt{matryoshka} emulator \citep{DonaldMcCann2021} trained on the BACCO simulations \citep{Angulo2021} reaches sub-percent precision up to $k\sim$ 1 h/Mpc.

Targeting non-linear scales also brings the difficulty from a more complex statistic: the matter/galaxy density field becomes significantly non-Gaussian, which increases the error bars due to non-Gaussian covariance terms. In this article, I build up on \cite{Lacasa2018b,Lacasa2019b} which allow a near complete modeling of these non-Gaussian terms using the halo model and Halo Occupation Distribution.

The question then rises of how much statistical power is indeed contained in these non-linear scales. And further, how much of this statistical power can effectively be harnessed for cosmological constraints. One could indeed intuitively think that most of this power would only constrain astrophysical/galaxy formation parameters. The purpose of this article is thus to investigate how much cosmological information, in particular on Dark Energy, is contained in the galaxy 2-point function depending on the range of scales of analysis, accounting both for the astrophysical dependence and the rising non-Gaussianity of the field. I will apply this specifically for the tomographic angular power spectrum, and for surveys with a high enough galaxy density that shot-noise is subdominant.

The modeling and analytical equations are described in section~\ref{Sect:HM-Clgal-Cov}, first stating the surveys I consider and the observational specifications in Sect.~\ref{Sect:setup}, then the halo modeling of the galaxy distribution in Sect.~\ref{Sect:HM}, followed in Sect.~\ref{Sect:cl_and_cov} by a recapitulation of the equations of non-Gaussian covariance terms and first results on the behaviour of the power spectrum and its covariance terms when pushing in the highly non-linear regime. In section~\ref{Sect:Fisher} I use Fisher forecasts to show the cosmological information content of the power spectrum as a function of scale, first in terms of multipoles in Sect.~\ref{Sect:angular-scales} and then translated into physical cuts $k_\mr{max}$ in Sect.~\ref{Sect:physical-scales}, before providing a physical interpretation of the results in Sect.~\ref{Sect:interpretation}. In section~\ref{Sect:higher-orders} I give an estimate of the information contained beyond the power spectrum and how it depends on scale. Finally I discuss the results and their potential consequences in section~\ref{Sect:discu}.

Throughout the article I adopt as fiducial cosmology flat $\Lambda$CDM with \textit{Planck} 2018 cosmological parameters \citep{Planck2018-cosmo}: $(\Omega_b h^2,\Omega_c h^2,H_0,n_S,\sigma_8,w_0)=(0.022,0.12,67,0.96,0.81,-1)$.

I make available at \url{https://github.com/fabienlacasa/SmallScaleMiracle} the data and Python notebooks that allow to reproduce all plots and results of the article, and some more.


\section{Halo modeling $\Clgal$ and its covariance}\label{Sect:HM-Clgal-Cov}

\subsection{Survey specifications and setup}\label{Sect:setup}

I consider two mock galaxy surveys for the forecasts: the \Euclid{} photometric sample and the \SKA{}2 galaxy survey where galaxies are detected as point sources in the HI intensity map.

For the \SKA{}2 sample I use specifications from \citep{Bull2016}: a sky coverage of 15'000 deg$^2$ and a galaxy number density given by Table~3 of \cite{Bull2016} that I interpolated at all necessary redshifts. The total density is $\sim$9 gals/arcmin$^{2}$ in the redshift range [0,2]. I divide the sample into 10 equi-populated redshift bins, finding that the corresponding bin stakes are $z={0.1,0.198,0.267,0.33,0.393,0.461,0.537,0.628,0.748,0.934,2.}$

For the \Euclid{} sample, I use specifications from \citep{Euclid-redbook,Euclid-IST}: a sky coverage $f_\mr{SKY}=0.36$, a galaxy number density
\ba\label{Eq:Ngal(z)-Euclid}
n(z) \propto  \left( \frac{z}{z_0} \right)^2 \exp{\left[ - \left( \frac{z}{z_0} \right)^{3/2} \right]} 
\ea
where $z_0 = z_{\rm m}/\sqrt{2}$ with $z_{\rm m}= 0.9$ the median redshift \citep{Euclid-redbook}. The total density is 30 gals/arcmin$^{2}$ in the redshift range [0,2.5]. Following \cite{Euclid-IST}, I divide the sample into 10 equi-populated redshift bins, whose bin stakes are $z={0.001,0.418,0.56,0.678,0.789,0.9,1.019,1.155,1.324,1.576,2.5}$.

In the article I will show only plots for the \SKA{}2 case, as the plots for the \Euclid{} sample are all qualitatively similar, and the scientific conclusions are the same. I make available online the plots for both surveys on \url{https://github.com/fabienlacasa/SmallScaleMiracle}.

For both surveys, the forecasts are produced with a binning of multipoles, as is customary in data analysis. Specifically, I define 50 bins spaced logarithmically in the range [30,50$\,$000]. The binning operator is then defined by
\be
P_{b,\ell} = 
\left\{
\begin{array}{ll}
	\frac{w_\ell}{w_{\ell_\mr{cen}} \, \Delta\ell} & \qquad \mathrm{if} \quad \ell \in \mr{bin}(b) \\
	0 & \qquad \mathrm{otherwise} \\
\end{array}
\right.
\ee
with $\ell_\mr{cen}$ the center of the multipole bin and $w_\ell$ a weighting scheme. Here I adopt the simple scheme $w_\ell = \ell$ which makes $w_\ell C_\ell$ roughly constant and thus improves the binning approximation.\\
The binned power spectrum and covariance are then given by
\ba
C_b &= P_{b,\ell} \; C_\ell \\
\Cov(C_b,C_{b'}) &= P_{b,\ell} \; P_{b',\ell'} \; \mathcal{C}_{\ell,\ell'}
\ea
I use these equations to bin the power spectrum and the Gaussian part of the covariance, which is diagonal. I have however found these equations to be too numerically intensive to be used for the non-Gaussian parts of the covariance when reaching tens of thousands of multipoles. Indeed, for 10 redshift bins, that would require computing $\mathcal{O}(10^{10-11})$ covariance elements before binning them. Instead, for these terms I have used the approximation that the correlation matrix varies smoothly within the bin so that the binned correlation can be approximated by the correlation at the central multipole $\ell_\mr{cen}$. I have checked that this approximation works to percent precision on parameter forecasts up to $\ell_\mr{max}=2000$, which is the limit where I could push the brute force computation. In the following, for simplicity binned quantities will be plotted at the center of the multipole bin. 

\subsection{Halo modeling}\label{Sect:HM}

I use the standard halo model, as reviewed for instance by \cite{Cooray2002}. In terms of ingredients, I use the halo mass function from \cite{Tinker2008} with the corresponding halo bias from \cite{Tinker2010}. The halo profile is the classic NFW profile \citep{Navarro1996}, with the concentration-mass relation from \cite{Bullock2001}. I model the distribution of galaxies using the Halo Occupation Distribution (HOD). Specifically I adopt one similar to \cite{Zehavi2011} : $N_\mr{gal}=N_\mr{cen}+N_\mr{sat}$, where the central galaxy follows a Bernoulli distribution with probability
\ba
P(N_\mr{cen}=1) = \frac{1}{2} \left(1 + \mr{Erf}\left(\frac{\log_{10}M-\log_{10}M_\mr{min}}{\sigma_{\mr{log}M}}\right)\right)
\ea
and the satellite galaxies follow a Poisson distribution for the satellite galaxies, conditioned to the presence of the central, with mean
\ba
\mathbb{E}\left[N_\mr{sat}|N_\mr{cen}=1\right]= \left(\frac{M}{M_\mr{sat}}\right)^{\alpha_\mr{sat}} .
\ea
In terms of spatial distribution, all galaxies are assumed to distribute stochastically following the halo profile.

The specifications for the galaxy redshift distribution $n(z)$ given in Sect.~\ref{Sect:setup} do not correspond to a volume-limited sample, i.e. constant HOD parameters. To deal with this, I follow the approach of \cite{Lacasa2019b} by fitting the HOD at each redshift. Specifically, I fit the $M_\mr{min}$ parameter, assuming that the ratio $M_\mr{ratio} = M_\mr{sat}/M_\mr{min}=10$ is constant and that $\sigma_{\log M}=0.5$ and $\alpha_\mr{sat}=1$ are constant. The resulting function $M_\mr{min}(z)$ is then further fitted with a fourth order polynomial:
\ba
M_\mr{min}(z) = M_\mr{min}^a + M_\mr{min}^b \, z + M_\mr{min}^c \, z^2 + M_\mr{min}^d \, z^3 .
\ea
\cite{Lacasa2019b} showed that this redshift-dependent gives a $\sim$2.5\% fit to $n(z)$ on the full redshift range and reproduces the galaxy bias from Euclid-internal simulations. I use the same approach for the \SKA{}2 sample, finding the same level of accuracy.
Specifically, I found the following values of the HOD parameters to best fit the specifications :
for \Euclid{} $M_\mr{min}^a=11.03$, $M_\mr{min}^b=-0.185$, $M_\mr{min}^c=0.575$, $M_\mr{min}^d=-0.107$ ;
for \SKA{}2, these were $M_\mr{min}^a=10.24$, $M_\mr{min}^b=2.53$, $M_\mr{min}^c=-0.610$, $M_\mr{min}^d=0.064$, $M_\mr{ratio}=10$, $\sigma_{\log M}=0.5$ and $\alpha_\mr{sat}=1$.
This approach enables a halo modeling of the galaxy sample over the whole redshift range with 7 parameters : $M_\mr{min}^a$, $M_\mr{min}^b$, $M_\mr{min}^c$, $M_\mr{min}^d$, $M_\mr{ratio}$, $\sigma_{\log M}$ and $\alpha_\mr{sat}$.

With this framework, all $n$-point polyspectra of galaxies can be computed through the halo model, and this involves integrals of the form:
\ba\label{Eq:unified-HM-integrals}
\nonumber I_\mu^\beta(k_1,\cdots,k_\mu|z) = \int \dd M \ & \frac{\dd n_h}{\dd M} \ \lbra N_\mr{gal}^{(\mu)}\rbra \ b_\beta(M,z) \\ 
& \times u(k_1|M,z) \cdots u(k_\mu|M,z)  
\ea
with $\frac{\dd n_h}{\dd M}$ the halo mass function, $u(k|M,z)$ the halo profile, $b_\beta(M,z)$ the halo bias or order $\beta$, and $\lbra N_\mr{gal}^{(n)}\rbra \equiv \lbra N_\mr{gal} (N_\mr{gal}-1)\cdots(N_\mr{gal}-(n-1))\rbra $ the number of $n$-uplets of galaxies (implicitely depending on halo mass). \\
For instance the number density of galaxies in a given redshift bin (in units galaxies/steradian) is given by
\be
N_\mr{gal}(i_z) = \int \dd V \ I_1^0(0|z)
\ee
where the integral runs implicitly over redshifts in the bin $i_z$, $\dd V = r^2(z) \frac{\dd r}{\dd z} \dd z$ is the comoving volume per steradian and $r(z)$ is the comoving distance to redshift z

\subsection{$\Clgal$ and its covariance}\label{Sect:cl_and_cov}

\begin{figure}[!ht]
\begin{center}
\includegraphics[width=\linewidth]{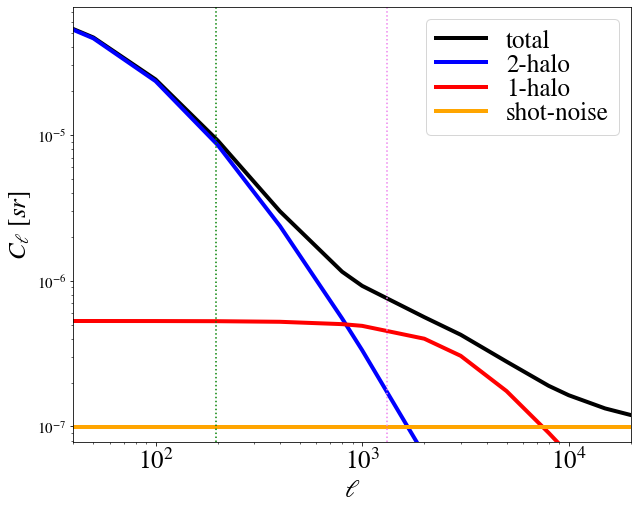}
\caption{Angular power spectrum of the \SKA{}2 galaxies in $z=0.46-0.53$, and its different terms. Green dotted vertical line : limit of perturbation theory k=0.15 h/Mpc ; violet dotted vertical line : limit of the \texttt{matryoshka} emulator k=1 h/Mpc.}
\label{Fig:Clgal}
\end{center}
\end{figure}

Using the halo model, the power spectrum is standardly composed of 2-halo, 1-halo and shot-noise terms:
\ba
C_\ell^\mr{2h}(i_z) =& \left.\int \dd V \ \left(I_1^1(k_\ell|z)\right)^2 P(k_\ell|z) \, \middle/ \ N_\mr{gal}(i_z)^2\right. \label{Eq:Cl2h} \\
C_\ell^\mr{1h}(i_z) =& \left.\int \dd V \ I_2^1(k_\ell,k_\ell|z) \, \middle/ \ N_\mr{gal}(i_z)^2\right. \label{Eq:Cl1h} \\
C_\ell^\mr{shot}(i_z) =& \left.\int \dd V \ I_1^0(0|z) \, \middle/ \ N_\mr{gal}(i_z)^2\right. = 1/N_\mr{gal}(i_z) \label{Eq:Clshot}
\ea
Fig.~\ref{Fig:Clgal} shows the resulting power spectrum for the case of \SKA{}2 galaxies in the redshift bin $z=0.46-0.53$, the median bin of the sample.

Two features are worth noting. First, the 1-halo term is roughly constant on multipoles $\ell \lesssim 2000$, but acquires a significant scale dependence afterwards. From Eq.~\ref{Eq:Cl1h} this scale dependence appears when we hit the radius of the typical host halo mass of the galaxy sample. Second, the high density of galaxies makes shot-noise subdominant, revealing the 1-halo term on a wide range of scales. Furthermore shot-noise can be subtracted exactly so what is important is its contribution to the covariance, where it contributes to the Gaussian part 
\be
\mathcal{C}_{\ell,\ell'}^G = \frac{2 \ \left(C_\ell^\mr{clust}(i_z)+C_\ell^\mr{shot}(i_z)\right)^2}{2\ell+1} \ \delta_{\ell,\ell'} \ \delta_{i_z,j_z}
\ee
With this I find that all multipoles of the power spectrum of \SKA{}2 galaxies can be measured with $(S/N)_G>5$ on the whole range $\ell\in [2,20\,000]$ for all redshift bins. Even higher significance is reached for the \Euclid{} photometric sample, which contains more galaxies. This shows that a huge statistical power will be present in the strongly non-linear regime, where the 1-halo dominates, both for \SKA{}2 and \Euclid{}.

The Gaussian formula however does not capture the full covariance, especially on small scales. In this article I follow the equations for the non-Gaussian part of the covariance from \cite{Lacasa2018b}, and the numerical approximation and implementation of \cite{Lacasa2019b}. For the article to be self-contained, I summarize here the involved terms. The non-Gaussian covariance is composed of different contributions: super-sample covariance (SSC), braiding covariance, 1-halo term, 2-halo 1+3 term, 3-halo base-0 term and 4-halo third order term.
\be
\mathcal{C}_{\ell,\ell'}^{NG} = \mathcal{C}_{\ell,\ell'}^{SSC} + \mathcal{C}_{\ell,\ell'}^{Braid} + \mathcal{C}_{\ell,\ell'}^{1h} + \mathcal{C}_{\ell,\ell'}^{2h1+3} + \mathcal{C}_{\ell,\ell'}^{3hbase0} + \mathcal{C}_{\ell,\ell'}^{4h3} 
\ee
The super-sample covariance is given by
\ba\label{Eq:SSC}
\nonumber \mathcal{C}_{\ell,\ell'}^\mr{SSC} = \int \dd V_{ab} \ \Psi_\ell^\mr{sqz}(z_a) \ \Psi_{\ell'}^\mr{sqz}(z_b) \ \sigma^2(z_a,z_b) \ \Bigg/ \ N_\mr{gal}(i_z)^2 \, N_\mr{gal}(j_z)^2 
\ea
where $z_a \in i_z$, $z_b \in j_z$,
\ba
\sigma^2(z_a,z_b) = \frac{C_{\ell=0}^m(z_a,z_b)}{4\pi} = \frac{1}{2\pi^2} \int k^2\,\dd k \ P(k|z_{ab}) \ j_0(k r_a) \ j_0(k r_b)
\ea
is the SSC kernel, and 
\ba
\Psi_\ell^\mr{sqz}(z) = 4 \ I_1^{\Sigma_2}(k_{\ell}|z) \ I_1^1(k_{\ell}|z) \ P(k_{\ell}|z) + I_2^1(k_{\ell},k_{\ell}|z)
\ea
where
\ba
I_\mu^{\Sigma_2} = \frac{17}{21} \ I_\mu^{1} + \frac{1}{2!} \ I_\mu^{2}
\ea
is the sum of contributions from second order perturbation theory and second order bias

For Braiding covariance I use the Bij approximation from \cite{Lacasa2019b}
\ba
\mathcal{C}_{\ell,\ell'}^\mr{Braid} = 2 \ \Psi^\mr{alt,int}_{\ell,\ell'}(i_z) \ \Psi^\mr{alt,int}_{\ell,\ell'}(j_z) \ B_{\ell,\ell'}(i_z,j_z)
\ea
where
\ba
\Psi^\mr{alt,int}_{\ell,\ell'}(i_z) = \int \dd V \ \Psi^\mr{alt}_{\ell,\ell'}(z)
\ea
and 
\be
\nonumber B_{\ell,\ell'}(i_z,j_z) = \sum_{\ell_a} \frac{2\ell_a+1}{4\pi} \  {\threeJz{\ell}{\ell'}{\ell_a}}^2 \ C_{\ell_a}^{n_g^2}(i_z,j_z)
\ee
with
\ba
C_{\ell}^{n_g^2}(i_z,j_z) = \int \dd V_{ab} \ \nbargal(z_a)^2 \, \nbargal(z_b)^2 \ C_{\ell}^{m}(i_z,j_z) \Bigg/ \left(I^{n_g^2}(i_z) \, I^{n_g^2}(j_z)\right)
\ea
and
\ba
I^{n_g^2}(i_z) = \int_{z\in i_z} \dd V \ \nbargal(z)^2 \ .
\ea
Then come the 1-halo term
\ba
\mathcal{C}_{\ell,\ell'}^\mr{1h} = \frac{\delta_{i_z,j_z}}{4\pi} \int \dd V \ I_4^0(k_{\ell},k_{\ell},k_{\ell'},k_{\ell'}|z) \ \Bigg/ \ N_\mr{gal}(i_z)^4 \ ,
\ea
the 2-halo 1+3 term
\ba
\nonumber \mathcal{C}_{\ell,\ell'}^\mr{2h1+3} = \frac{2\delta_{i_z,j_z}}{4\pi} & \int \dd V \ I_1^1(k_\ell|z) \ I_3^1(k_\ell,k_{\ell'},k_{\ell'}|z) \ P(k_\ell|z) \ \Bigg/ \ N_\mr{gal}(i_z)^4 \\
& + (\ell \leftrightarrow \ell') \ ,
\ea
the 3-halo base term
\ba
\nonumber \mathcal{C}_{\ell,\ell'}^\mr{3h-base0} = \frac{\delta_{i_z,j_z}}{4\pi} & \int \dd V \ 2 \;  \left(I_1^{1}(k_\ell|z) \ P(k_\ell|z)\right)^2 I_2^{\Sigma_2}(k_{\ell'},k_{\ell'}|z) \Bigg/ N_\mr{gal}(i_z)^4 \\
\nonumber & + \quad (\ell\leftrightarrow\ell') \\
\nonumber +\frac{4 \ \delta_{i_z,j_z}}{4\pi} & \int \dd V \ 2 \ I_1^{1}(k_\ell|z) \ I_1^{1}(k_{\ell'}|z) \ I_2^{\Sigma_2}(k_{\ell},k_{\ell'}|z) \\
& \times P(k_\ell|z) \ P(k_{\ell'}|z) \ \Bigg/ \ N_\mr{gal}(i_z)^4 \ ,
\ea
and the 4-halo term from third order contributions
\ba
\nonumber \mathcal{C}_{\ell,\ell'}^\mr{4h-3} = & \frac{2 \ \delta_{i_z,j_z}}{4\pi} \int \dd V \ 3! \ \left(I_1^1(k_{\ell},z)\right)^2 \ I_1^1(k_{\ell'},z) \ I_1^{\Sigma_3}(k_{\ell'},z) \\
& \times \ P(k_{\ell}|z) \ P(k_{\ell}|z) \ P(k_{\ell'}|z) \Bigg/ \ N_\mr{gal}(i_z)^4 \quad + \quad (\ell\leftrightarrow\ell') \ .
\ea
where
\ba
I_\mu^{\Sigma_3} \equiv \frac{1023}{1701} \ I_\mu^{1} + \frac{1}{3!} \ I_\mu^{3}
\ea
is the sum of contributions from third order perturbation theory and third order bias.

\begin{figure}[!ht]
\begin{center}
\includegraphics[width=\linewidth]{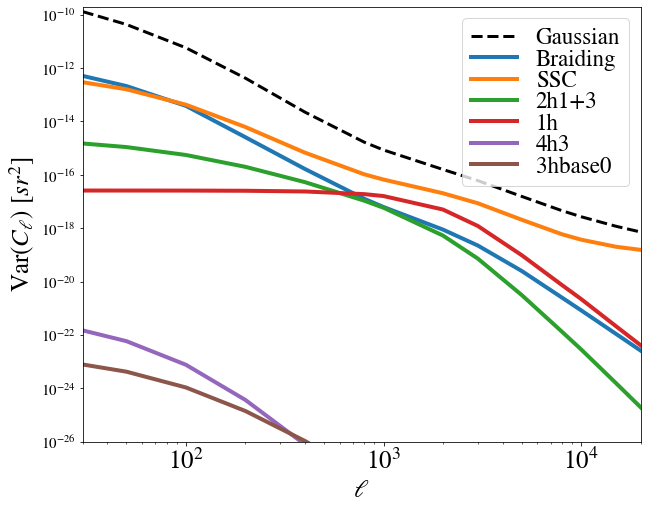}
\caption{Different contributions to the variance of \Clgal{} per individual multipole, for \SKA{}2 galaxies in the median bin $z=0.46-0.53$.}
\label{Fig:Var-Cl}
\end{center}
\end{figure}

Figure~\ref{Fig:Var-Cl} shows all these terms in the case of the variance $\mathcal{C}_{\ell,\ell}$ in the median bin of \SKA{}2.

We see that the 3h-base and 4h-3 terms are negligible for this variance. In fact \cite{Lacasa2019b} already found that they have a negligible impact on the total covariance as well as parameter constraints up to $\ell_\mr{max}=2000$. Here I find this result to still hold to $\ell_\mr{max}=20\,000$.

\begin{figure}[!ht]
\begin{center}
\includegraphics[width=\linewidth]{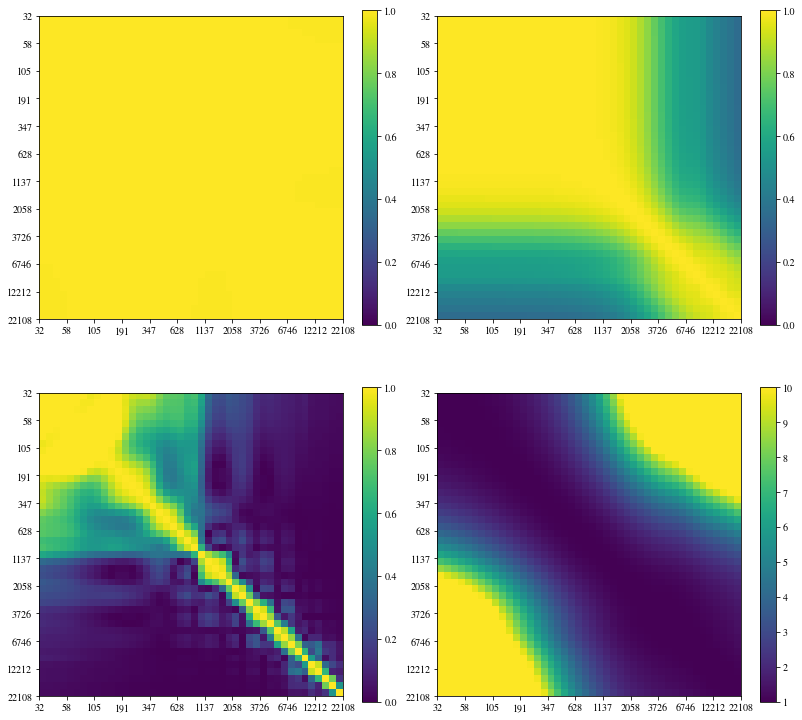}
\caption{Correlation matrices for the different non-Gaussian covariance terms, normalised by their own diagonal, for \SKA{}2 galaxies in $z=0.46-0.53$. \textit{From left to right and top to bottom:} SSC, 1-halo, Braiding and 2-halo 1+3.}
\label{Fig:Corr-Cl}
\end{center}
\end{figure}

Figure~\ref{Fig:Var-Cl} however does not let us appreciate the complexity and importance of the other non-Gaussian terms, because it only focuses on the diagonal multipole by multipole. Also real analyses bin multipoles together, which can significantly change the amplitude of the terms ; for instance the Gaussian variance decreases strongly (typically as $1/\Delta\ell$). Thus all later results apply the binning of multipoles explained in Sect.~\ref{Sect:setup}. So, after binning, for the four most important non-Gaussian terms, Figure~\ref{Fig:Corr-Cl} shows the correlation matrix $\mathcal{C}_{\ell,\ell'}/\sqrt{\mathcal{C}_{\ell,\ell} \times \mathcal{C}_{\ell',\ell'}}$, where each term is normalised by its \emph{own} diagonal, to reveal its specific structure.

We see results consistent with \cite{Lacasa2019b} on large scales: SSC and 1-halo both yield 100\% correlated covariance, Braiding is also strongly correlated albeit lower, and 2-halo 1+3 is minimal on the diagonal as it correlates large scales with small scales. On top of this, a new behaviour appears at $\ell \gtrsim 2000$ : the 1-halo stops being 100\% correlated and gets closer to diagonal ; the same behaviour happens for Braiding covariance.

\begin{figure}[!ht]
\begin{center}
\includegraphics[width=\linewidth]{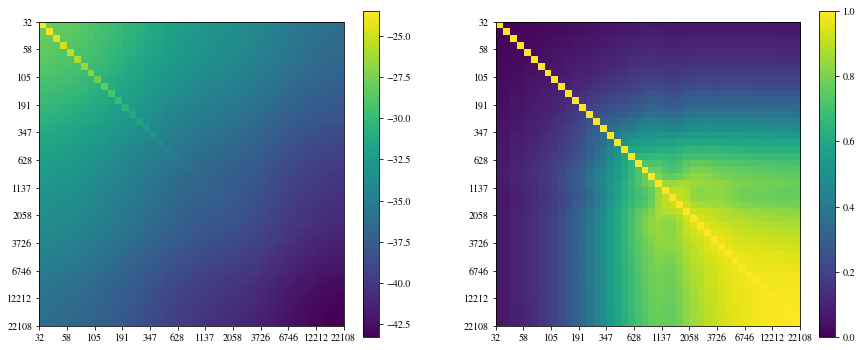}
\caption{Total covariance matrix, including both Gaussian and non-Gaussian terms, for \SKA{}2 galaxies in $z=0.46-0.53$. \textit{Left:} covariance in log color scale. \textit{Right:} correlation matrix.}
\label{Fig:Cov-Cl}
\end{center}
\end{figure}

Now to see a first estimation of the relevance of these non-Gaussian terms, Fig.~\ref{Fig:Cov-Cl} shows the total covariance, including both the Gaussian and non-Gaussian contributions.

Remembering that the Gaussian covariance only contributes on the diagonal, we see that the non-Gaussian term become relevant already at multipoles of a few hundreds, and at multipoles of a few thousands they dominate the matrix to the point that it becomes >90\% correlated.


\section{Fisher constraints}\label{Sect:Fisher}

In this section, I use Fisher forecasts to quantify the information content in the galaxy angular power spectrum, and how it varies when extending the range of multipole of analysis. I use the covariance matrices shown in the previous section, rescaled with the $f_\mr{SKY}$ approximation
\be
\mathcal{C}_\mr{partial-sky} = \left.\mathcal{C}_\mr{full-sky} \middle/ f_\mr{SKY}\right.
\ee
to account for the partial sky coverage of the surveys. The Fisher information matrix in a given redshift bin then follows:
\ba
F_{\alpha,\beta}(i_z) = \sum_{\ell,\ell'=\ell_\mr{min}}^{\ell_\mr{max}} \partial_\alpha C_\ell^\mr{gal}(i_z) \ \mathcal{C}^{-1}_{\ell,\ell'}(i_z,i_z) \ \partial_\beta C_{\ell'}^\mr{gal}(i_z)
\ea
and the matrix summed over all bins is:
\ba
F_{\alpha,\beta} = \sum_{i_z,j_z} \sum_{\ell,\ell'=\ell_\mr{min}}^{\ell_\mr{max}} \partial_\alpha C_\ell^\mr{gal}(i_z) \ \mathcal{C}^{-1}_{\ell,\ell'}(i_z,j_z) \ \partial_\beta C_{\ell'}^\mr{gal}(j_z)
\ea
where $\alpha,\beta$ are model parameters, i.e. in the following both cosmological parameters $(\Omega_b h^2, \Omega_c h^2, H_0, n_S, \sigma_8, w_0)$ and HOD parameters $(\alpha_\mr{sat}, \sigma_{\log M}, M_\mr{ratio}, M_\mr{min}^a, M_\mr{min}^b, M_\mr{min}^c, M_\mr{min}^d)$. $\partial_\alpha C_{\ell}^\mr{gal}$ is the derivative of the power spectrum with respect to the parameter $\alpha$, and for simplicity I denote multipole bins with their center in all the following.

\subsection{In angular scales}\label{Sect:angular-scales}

\begin{figure}[!ht]
	\begin{center}
		\includegraphics[width=\linewidth]{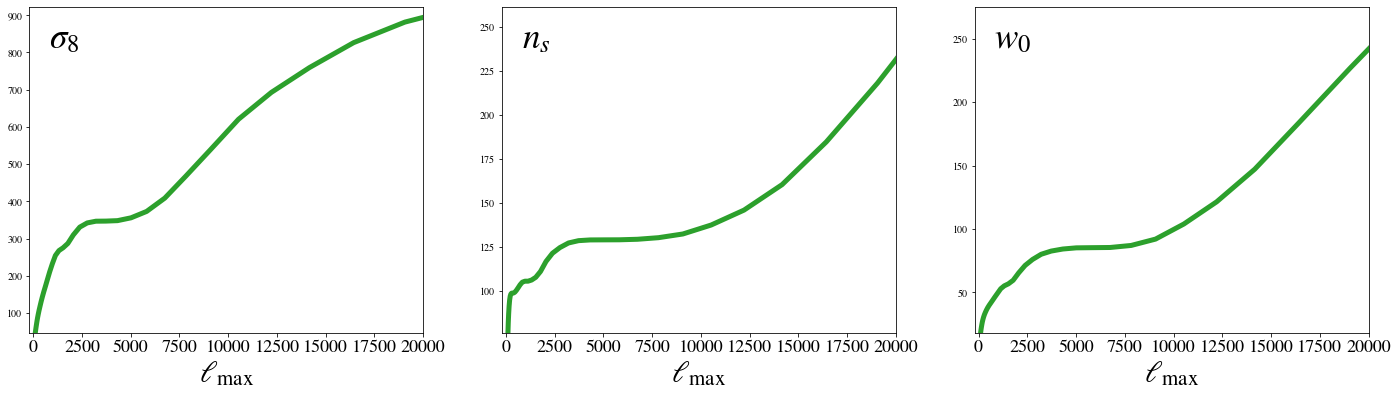}
		\caption{Fisher information $F_{\alpha,\alpha}^{1/2}$ on $\sigma_8$, $n_S$ and $w_0$ as a function of the angular cut $\lmax$, for \SKA{}2 galaxies in $z=0.46-0.53$.}
		\label{Fig:Fish-sig8nsw-lmax}
	\end{center}
\end{figure}

First, keeping $\ell_\mr{min}$ fixed, I study how the Fisher information varies when increasing the maximum multipole of analysis $\ell_\mr{max}$. For this, I concentrate on the case of the three cosmological parameters that can best be constrained with full shape galaxy power spectra : $\sigma_8$, $n_S$ and $w_0$. In the case of the median redshift bin of \SKA{}2,  Figure~\ref{Fig:Fish-sig8nsw-lmax} shows the information $F_{\alpha,\alpha}^{1/2}$ as a function of $\ell_\mr{max}$. This quantity is indeed the inverse of the error bar on the parameter $\alpha$ before marginalisation on other model parameters.

We see a typical striking behaviour of the curves : when increasing $\ell_\mr{max}$ the information first rises steadily on linear and weakly non-linear scales, but this increase progressively slows down before coming to a stop around $\lmax\sim2000$, a plateau is then present where information has saturated : adding these power spectrum measurements does not bring any (direct) information on cosmological parameters. The extension of this plateau depends on the considered parameter, but for all of them the plateau finally comes to an end and information rises again with a steep slope. The information brought after the plateau up to $\lmax=20\,000$ is comparable to (for $n_S$) or larger than (for $\sigma_8$ and $w_0$) the information from linear/weakly non-linear scales before the plateau. This steep rise is the small scale miracle that gives its title to this article. 

\begin{figure}[!ht]
	\begin{center}
		\includegraphics[width=.49\linewidth]{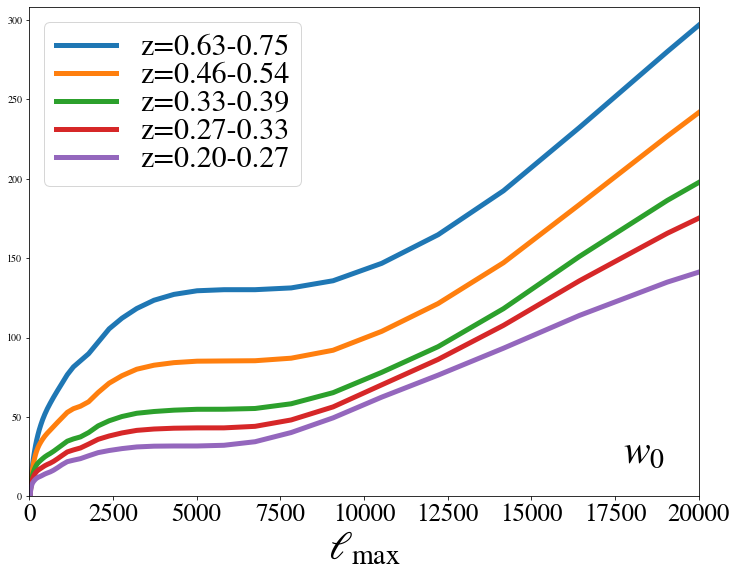}
		\includegraphics[width=.49\linewidth]{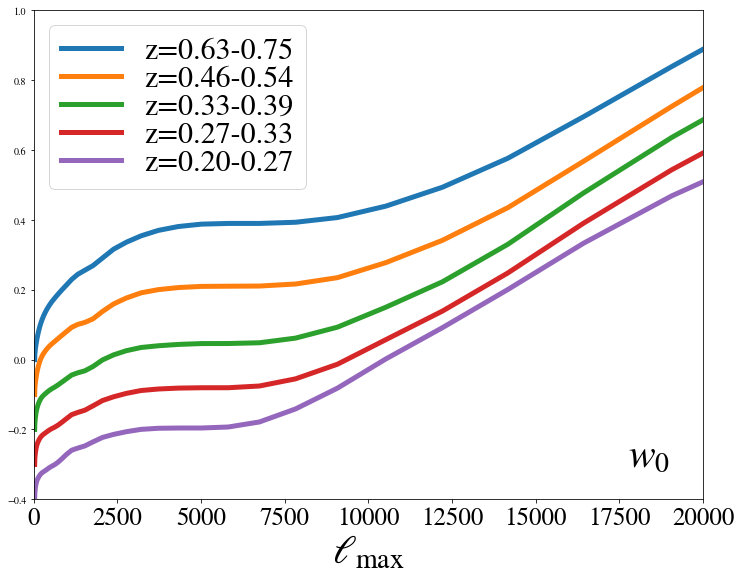}
		\caption{\textit{Left:} Fisher information on $w_0$, $F_{w_0,w_0}^{1/2}$, as a function of $\lmax$ for most redshift bins of \SKA{}2, where the bins were selected to keep clarity of the plot. \textit{Right:} same but all curves are rescaled to [0,1] and an offset is added for readability.}
		\label{Fig:Fish-w-lmax}
	\end{center}
\end{figure}

Now comes the question whether this qualitative behaviour is peculiar to this redshift bin and this survey. Concentrating on the case of Dark Energy, Figure~\ref{Fig:Fish-w-lmax} shows $F_{w_0,w_0}^{1/2}$ as a function of $\lmax$ for different redshift bins of \SKA{}2, both raw (left) and after a rescaling to appreciate the qualitative behaviour of each curve (right). I selected 5 out of the 10 redshift bins to avoid overcrowding the plot, but the omitted bins follow the same qualitative behaviour as the one presented.

We see that all curves follow the qualitative behaviour found previously, with a striking plateau where information saturates before rising again, the ``small scale miracle''. The location (and extent) of the plateau depends on redshift : it slowly moves to smaller angular scales at higher redshifts. I checked that this qualitative behaviour of the information content is also present for the \Euclid{} survey, the main difference being that the plateau can move to even smaller scales as the sample goes to higher redshift (up to $\ell\sim12\,000$ in the highest bin $z=1.6-2.5$). I also checked whether this qualitative behaviour is also present for other model parameters, and found that it is also clearly present for all the HOD parameters ; for the other cosmological parameters $\Omega_b h^2$ and $\Omega_c h^2$ ($H_0$ being basically unconstrained by the galaxy power spectrum) the picture is a bit less clear : a plateau can be seen in some redshift bins but not all, however the rise of information on small scales remains, which is the most important feature.

Before interpreting these results, we may wonder what these angular scales of the plateau correspond to in terms of physical scales, and whether the redshift evolution of the plateau could be explained by projection onto angles.

\subsection{In physical scales}\label{Sect:physical-scales}

In this section, I want to define a cut-off of the power spectrum data vector in physical scales $\kmax$ instead of angular scales $\lmax$. This for two reasons : (i) it allows more natural and redshift-independent understanding of the considered scales and what they correspond to, and (ii) the range of validity of theoretical approaches to non-linearity (perturbation theory, halo model, emulators...) is usually stated in terms of physical scales. 

A fixed cut-off in physical scales $\kmax$ corresponds to a redshift-dependent cut-off in angular scales $\lmax(z)=\kmax/r(z)$ with the Limber approximation. That prescription works well for infinitesimal redshift bins and when all multipoles are available. However in the present case, there is the added complexity that I have predefined redshift and multipole bins. I chose to apply the prescription at the center of the redshift bin to obtain a first $\lmax$, and then I cut at the multipole bin whose center is the closest to that $\lmax$. 

I defined 16 values of $\kmax$ between 0.1 Mpc$^{-1}$ --cut-off of the validity of perturbation theory-- to 10 Mpc$^{-1}$ --optimistic cut-off assuming future advances over the current emulators. In the lowest redshift bin, this can correspond to $\lmax$ above 20$\,$000, hence why I defined multipole bins up to higher $\ell$ in Sect.~\ref{Sect:setup}.

\begin{figure}[!ht]
	\begin{center}
		\includegraphics[width=\linewidth]{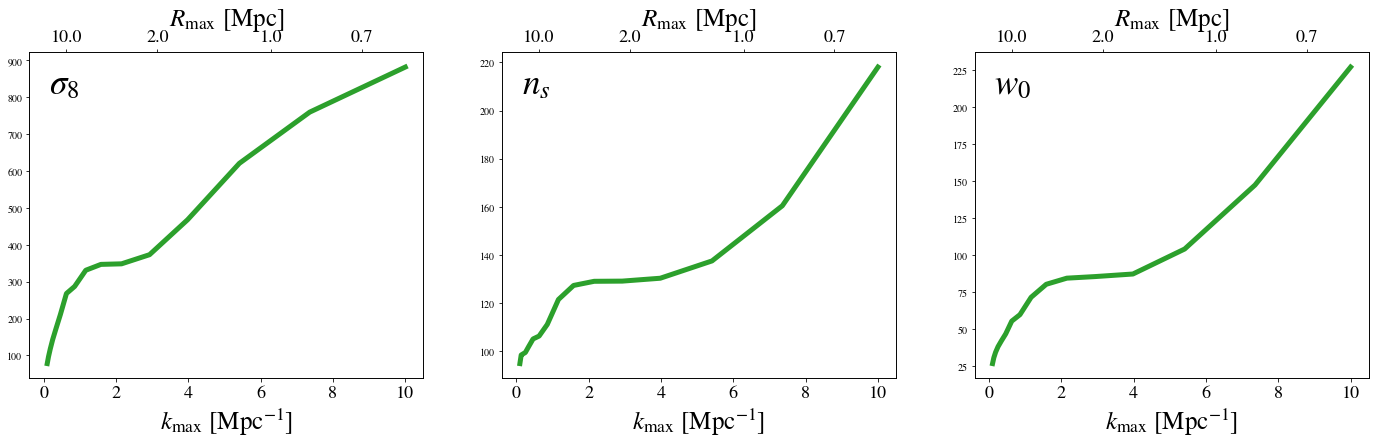}
		\caption{Fisher information on $\sigma_8$, $n_S$ and $w_0$ as a function of the physical cut $\kmax$, for \SKA{}2 galaxies in $z=0.46-0.53$. The corresponding real-space cut $R_\mr{max} = 2\pi/\kmax$ is indicated in the upper x-axis.}
		\label{Fig:Fish-sig8nsw-kmax}
	\end{center}
\end{figure}

Figure~\ref{Fig:Fish-sig8nsw-kmax} shows the resulting information content $F_{\alpha,\alpha}^{1/2}$ as a function of $\kmax$ for the cosmological parameters $\sigma_8$, $n_S$ and $w_0$, in the case of the median redshift bin of \SKA{}2.

We see that the typical behaviour found in Sect.~\ref{Sect:angular-scales}, with a plateau of information and the ``small scale miracle'', is still present. Furthermore, the plateau happens at a few Mpc$^{-1}$, which translates in real space to $\sim$2 Mpc. This is an interesting result as it basically corresponds to the limit of validity of current emulators.

\begin{figure}[!ht]
	\begin{center}
		\includegraphics[width=.49\linewidth]{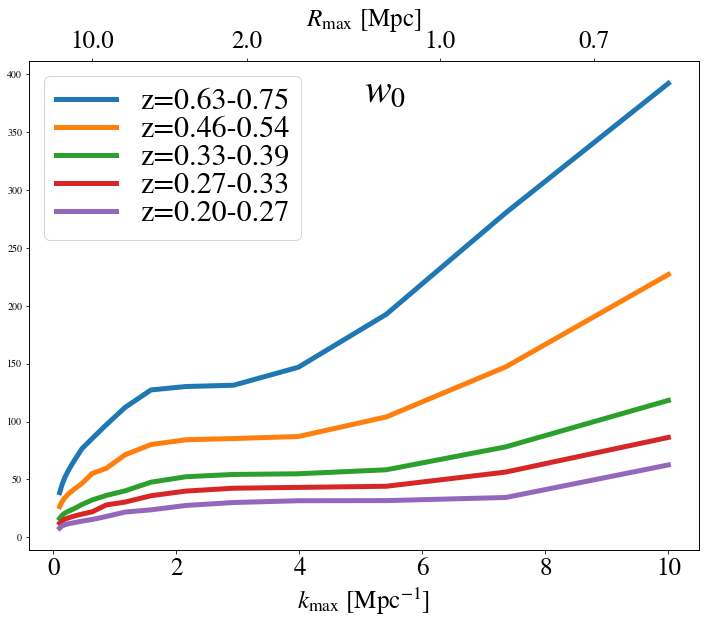}
		\includegraphics[width=.49\linewidth]{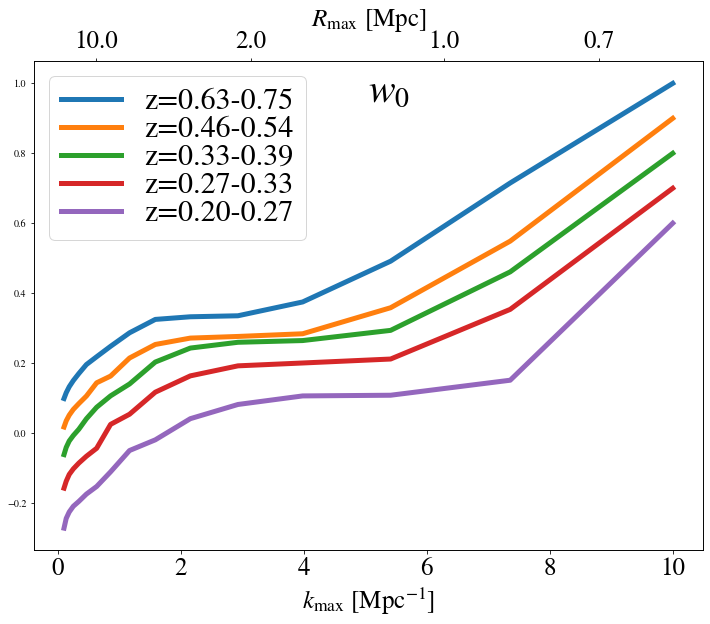}
		\caption{\textit{Left:} Fisher information on $w_0$ as a function of $\kmax$, for most redshift bins of \SKA{}2. \textit{Right:} same but all curves are rescaled to [0,1] and an offset is added for readability. The curves are not always perfectly smooth because the translation of $\kmax$ into $\lmax$ is approximate due to binning, as explained in the text.}
		\label{Fig:Fish-w-kmax}
	\end{center}
\end{figure}

Then, concentrating on the case of Dark Energy, Figure~\ref{Fig:Fish-w-kmax} shows the information content on $w_0$ as a function of $\kmax$ for different redshift bins of \SKA{}2.

We see that the location (and extent) of the plateau is redshift dependent, moving to larger physical scales when increasing redshift. In fact, this explains why the angular shift of the plateau in Figure~\ref{Fig:Fish-w-lmax} was only weak : when increasing redshift, a fixed physical scale projects onto smaller angular scales, so both effects (physical mode and projection) go in opposite directions.

\subsection{Interpretation}\label{Sect:interpretation}

I now give a physical understanding for the presence and location of the plateau of information and the subsequent rise of information on highly non-linear scales.

First, we have seen that the plateau appears around $\ell\sim2000$. Looking at Fig.~\ref{Fig:Clgal}, we see that this is a scale where the 1-halo term dominates the power spectrum and is still roughly constant. Looking at Fig.~\ref{Fig:Var-Cl} we see that the non-Gaussian covariance is dominated by the SSC and 1-halo at this scale. And looking at Fig.~\ref{Fig:Corr-Cl} we see that these covariance terms are 100\% correlated at this scale. So a first rough statistical picture is that on the scales of the plateau, we are measuring a (roughly) constant power spectrum whose error bars are 100\% correlated. So adding scales does not refine the measurement of that constant, and hence the information content saturates.

From a more physical point of view, the 1-halo power spectrum corresponds on large scales to the shot-noise (sometimes called Poisson noise) of the halos. The large scale value of $C_\ell^{1h}$ is a weighted average of the halo number (divided by $N_\mr{gal}(i_z)^2$), where massive halos have the more weight as they host more galaxies. So basically on large scales the 1-halo measures a single information : a weighted number of halos in the survey. At $\ell\sim2000$ we have basically exhausted the constraining power of this information. Indeed there is a cosmic variance to the number of halos, which is given by the 1-halo trispectrum (which quantifies the variance due to the discreteness of halos / Poisson noise) and super-sample covariance (which quantifies the variance due to super-survey fluctuations which modulate the number of halos inside the survey). Once this constraining power has been exhausted, as long as $C_\ell^{1h}$ keeps (roughly) constant, adding scales does not bring anything new. Hence the information content saturates.

Second, we can now understand why this plateau ends and information rises again. The start of this rise depends on the cosmological parameter but is roughly around $\ell\sim5000$. From Fig.~\ref{Fig:Clgal}, this is a scale where the $C_\ell^{1h}$ picks up a significant scale dependence. From Fig.~\ref{Fig:Var-Cl} and \ref{Fig:Corr-Cl}, on these scales the SSC dominates the covariance and is still 100\% correlated, while other sources of covariance become closer to diagonal. From a statistical point of view we are thus now able to extract information from the scale dependence of the power spectrum, and we are thus recovering constraining power. From a physical point of view, on these smaller scales $C_\ell^{1h}$ stops being dominated by massive (and rare) halos and starts to probe the amount of less massive halos which are smaller ; we no more measure a single information and thus recover constraining power.

Finally, we can refine the picture to understand why the location and extent of the plateau depends on cosmological parameter and redshift.\\
First for cosmological parameters, the cosmological constraints in the highly non-linear regime come from the fact that by pushing to smaller scales we measure a weighted number of halos with different weights ; so basically we measure the halo mass function (which is sensitive to cosmology), with smaller scales allowing to probe smaller masses. The halo mass function is highly sensitive to $\sigma_8$ with a steep scaling at high mass, and the exponent of that scaling decreases significantly at lower masses which allows to quickly break degeneracies. This explains why the plateau of information for $\sigma_8$ ends the soonest in Fig.~\ref{Fig:Fish-sig8nsw-lmax} : as soon as $C_\ell^{1h}$ stops being constant, its shape is violently sensitive to $\sigma_8$ due to the high scaling ; furthermore this scaling is completely different from that of the linear $C_\ell^{2h}$ which goes as $\sigma_8^2$. By contrast, $n_S$ shows a much longer plateau in Fig.~\ref{Fig:Fish-sig8nsw-lmax} ; this is because what we need to constrain $n_S$ is a high leverage arm in the scales $k$ of the initial power spectrum. The halo mass function is more weakly sensitive to $n_S$ (compared to $\sigma_8$), because what enters its prediction is $\sigma(R)$ the variance of the matter field smoothed at the halo radius scale, which is a convolution of $P(k)$ with a wide kernel. One needs to go to very small masses to probe small $k$ in $P(k)$. This explains why in Fig.~\ref{Fig:Fish-sig8nsw-lmax} the information on $n_S$ starts to rise again only at smaller scales, and why the ratio of information after/before the plateau is more modest for $n_S$ than $\sigma_8$. For the Dark Energy equation of state, the situation is intermediate between $\sigma_8$ and $n_S$ ; this is because the halo mass function is sensitive to $w_0$, though not as much as $\sigma_8$, and $w_0$ also has an influence on the comoving volume $\dd V$ which enters $C_\ell^{1h}$.\\
Second for the redshift dependence, we see from  Fig.~\ref{Fig:Fish-w-kmax} that the plateau extends to smaller scales at lower redshifts. This is basically due to sample selection : because the \Euclid{} and \SKA{} surveys are flux-limited (instead of volume-limited), at lower redshifts we have many more faint galaxies that live in light halos. So at low z $C_\ell^{1h}$ gets more contribution from less massive halos and will thus pick up a significant scale dependence only on smaller physical scales. This leads the plateau to extend to these smaller physical scales.

\subsection{Marginalisation over astrophysics}\label{Sect:marginalisation}

\begin{figure}[!ht]
	\begin{center}
		\includegraphics[width=\linewidth]{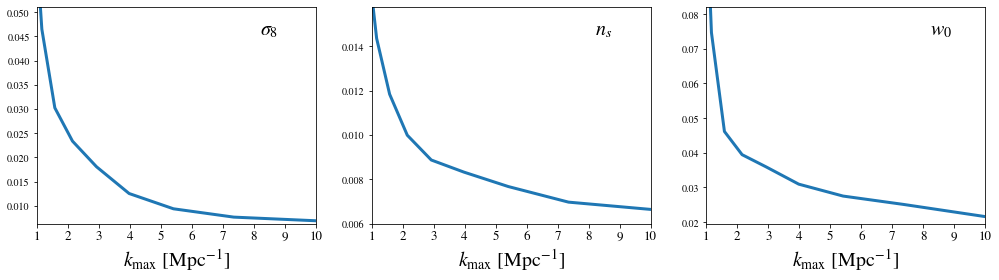}
		\caption{Marginalised error bars on $\sigma_8$, $n_S$ and $w_0$ as a function of $\kmax$, for \SKA{}2 galaxies summed over all redshift bins.}
		\label{Fig:marginalised-errors-kmax}
	\end{center}
\end{figure}

One issue we can worry about, is whether the cosmological information found on small scale is or not degenerate with the astrophysics. Figure~\ref{Fig:marginalised-errors-kmax} thus shows the marginalised error bars on $\sigma_8$, $n_S$ and $w_0$ as a function of $\kmax$, for \SKA{}2 galaxies summed over all redshift bins.

We see that, after marginalising over 7 HOD parameters there is still a substantial improvement of the cosmological errors when increasing $\kmax$. Quantitatively, when pushing $\kmax$ from 1 Mpc$^{-1}$ to 10 Mpc$^{-1}$, the error bar on $\sigma_8$ improves by a factor 9, the error bar on $n_s$ improves by a factor 2.5, and the error bar $w_0$ improves by a factor 5.2.\\
In comparison, pushing $\kmax$ from 0.1 Mpc$^{-1}$ to 1 Mpc$^{-1}$ yields larger improvements ratios, respectively a factor 25 for $\sigma_8$, 36 for $n_s$ and 11 for $w_0$. So the astrophysical uncertainties did hamper the improvements of cosmological constraints more in the small scale case ($1\rightarrow 10$ Mpc$^{-1}$) than in the large scale case ($0.1\rightarrow 1$ Mpc$^{-1}$). However the resulting improvements still seem worth the effort.

\subsection{Potential caveats}\label{Sect:caveats}

Many additional uncertainties in the galaxy modeling could hamper the extraction of the cosmological information. It is beyond the scope of this article to study them in detail. It is nonetheless interesting and contextualising to state them. These uncertainties can be classified in three categories: uncertainties on dark matter halo properties, on the distribution of galaxies inside halos, and on the impact of baryons.

First, properties of dark matter halos may be more complex and uncertain than assumed here. For instance, at the precision of future surveys, it was found that uncertainties in the fitting parameters of the halo mass function should be accounted for \citep{Artis2021}. The halo bias may also not only depend on their mass but other halo parameters (e.g. age, concentration, spin...) as well, an effect called assembly bias \citep{Shi2018,Sato-Polito2019} that would weaken how the large scale galaxy bias constrains HOD parameters. Furthermore, there is uncertainties in the halo shape: scatter should be included in the concentration--mass relation, although this seems to have a negligible impact on the information content \citep{Rizzato2019}, and more generally one should go beyond the usual assumption of spherical cows e.g. by accounting for triaxiality \citep{Smith2005}.

Second, the distribution of galaxies inside halos can have more uncertainties. More redshift dependence may be allowed for the HOD parameters, notably the high mass slope $\alpha_\mr{sat}$ and the ratio $M_\mr{sat}/M_\mr{min}$. The stochastic number of galaxies inside a halo may not follow a Poisson distribution \citep{Cacciato2013}. Furthermore, one should consider that central galaxies have a different spatial distribution than satellites, with a possibility of off-centering for some of them \citep{More2015}. Generally, galaxies could be allowed to follow a profile different from the dark matter NFW profile, though current studies indicate this is not necessary \citep{Bose2019}.

Third, the impact of baryonic physics, and the uncertainties this generates, is currently a hot topic. It is long known that baryons impact significantly the matter power spectrum on small scales \citep{Jing2006,Levine2006}, leading to changes of order 10-30\% on the scales relevant to this article \citep{vanDaalen2011}. But baryons also impact the halo properties: the halo mass function \citep{Cui2012}, the distribution of sub-halos \citep{RomanoDiaz2010} and halo profiles \citep{Abadi2010,Duffy2010}. Finally they also impact the HOD of hosted galaxies \citep{Bose2019,Beltz-Mohrmann2020}, and the spatial distribution of galaxies inside halos \citep{vanDaalen2014}.

The impact of a restricted number of these additional uncertainties is studied in appendix \ref{Sect:ext-astro}, with an extended model that allows for more redshift dependence of the HOD, deviation of the Poisson law for the galaxy distribution, and more freedom in their spatial distribution. This is not meant to be an exhaustive analysis. It shows that the cosmological constraints are indeed degraded, though the degradation is under reasonable control on small scales. At scales $k_\mr{max} \lesssim 1-2$ Mpc$^{-1}$, the current priors on these additional parameters are nearly enough to mitigate their introduction. On smaller scales, these parameters only degrade constraints on $n_S$, and better priors would be helpful.


\section{Estimating the information in higher orders}\label{Sect:higher-orders}

Until now I have focussed the study on the information contained solely in the power spectrum. A first question that arises is whether the qualitative behaviour that I have found, with the plateau and the ``small scale miracle'', can be expected for higher order statistics. Indeed, future surveys do plan to use this information, for instance at third order with the bispectrum.\\
From analytical arguments, I indeed expect a similar behaviour for higher order correlation functions / polyspectra. Indeed these polyspectra also get dominated by a 1-halo term in the highly non-linear regime. This 1-halo polyspectrum will also be constant on large scales and pick up a scale dependence at the radius corresponding to the typical host halo mass. So before that scale, the 1h information will be a constant equal to a weighted number of halo in the survey (with weights different from the power spectrum, preferring even more massive halos), and that constant will have a cosmic variance due to SSC and 1h covariance which will limit its constraining power.

\begin{figure}[!ht]
	\begin{center}
		\includegraphics[width=\linewidth]{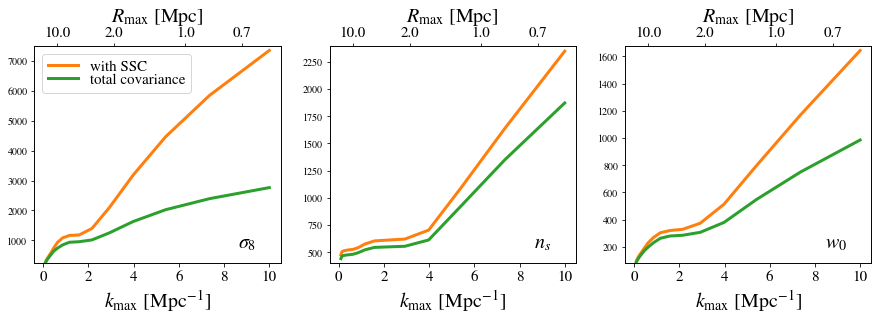}
		\caption{Fisher information on $\sigma_8$, $n_S$ and $w_0$ as a function of the physical cut $\kmax$, for \SKA{}2 galaxies summed over all redshift bins. The green lower curve shows the power spectrum information with the total covariance, while the orange upper curve uses only the Gaussian + super-sample covariance, as an estimate of the fundamental limit of the information in the galaxy density field.}
		\label{Fig:Fish-allz-sig8nsw-kmax}
	\end{center}
\end{figure}

A second question that arises is how much constraining power these higher orders can bring on top of the power spectrum. In other words, what is the fundamental limit to the constraining power of the galaxy density field, if we knew how to analyse it optimally ?\\
If the galaxy density field were Gaussian, the power spectrum would be the optimal statistic, so the limit would be given by the Fisher information in the power spectrum with a Gaussian covariance. The non-linearity however introduces another fundamental limit : super-sample covariance. Indeed, SSC comes from super-survey modes which change the matter density in the survey by an amount $\delta_b$ called the background shift. \cite{Wagner2015} has shown that a portion of Universe with this background shift evolves identically to a portion of Universe with a different cosmology. This is the basis of the so-called separate universe approach. Observationally, this means that all cosmological observables, and in particular all statistics of the galaxy density field, will behave as in this different cosmology. Super-sample covariance thus sets a fundamental limit to the cosmological constraints achievable from a given survey volume, independently of the statistics used. Figure~\ref{Fig:Fish-allz-sig8nsw-kmax} hence compares the Fisher information with the Gaussian+SSC covariance on one hand and the total power spectrum covariance on the other hand, in the case of constraints on $\sigma_8$, $n_S$ and $w_0$ summed over all redshift bins of \SKA{}2.

We see that there is indeed significant information that can be gained beyond the power spectrum, and that this gain rises rapidly when pushing to smaller scales. To be more quantitative, Table~\ref{Table:increase-Fisher-HighOrder} gives the increase of information on $\sigma_8$, $n_S$ and $w_0$ for some scale cuts.

\begin{table}
	\begin{center}
		\begin{tabular}{c|c|c|c|c} 			
			& 1 Mpc$^{-1}$ & 2 Mpc$^{-1}$ & 5 Mpc$^{-1}$ & 10 Mpc$^{-1}$ \\
			\hline
			$\sigma_8$ & +26\% & +32\% & +114\% & +166\% \\
			\hline
			$n_S$ & +9.5\% & +11\% & +17\% & +25\% \\
			\hline
			$w_0$ & +16\% & +15\% & +43\% & +67\% \\
		\end{tabular}
	\end{center}
	\caption{Improvement of the Fisher information on $\sigma_8$, $n_S$ and $w_0$ between the Gaussian+SSC and the total covariance, depending on the scale cut in the analysis, using all redshift bins of \SKA{}2.}\label{Table:increase-Fisher-HighOrder}
\end{table}

The gain is the most spectacular for $\sigma_8$, more modest but still interesting for $n_S$, and important for $w_0$. This can be understood because $\sigma_8$ highly influences the amount of non-linearity and thus high order statistics allow excellent constraints by breaking degeneracies ; for instance on perturbative scales the power spectrum scales as $(b \sigma_8)^2$ while the bispectrum scales as $b^3 \sigma_8^4$, where b is the galaxy bias. For $n_S$, as argued in Sect.~\ref{Sect:interpretation} constraints in the highly non-linear regime come from probing small mass halos ; however high order statistics preferentially probe extreme events i.e. massive halos, so the improvement they bring is more modest. Finally for $w_0$ the situation is intermediate between $n_S$ and $\sigma_8$ because it impacts the mass function at all masses, and because it also gets constrained by the comoving volume $\dd V$ whose constraints are improved by high orders. Indeed, for the numerator of the power spectrum (i.e. without the $1/N_\mr{gal}(i_z)^2$ normalisation), the 2-halo term scales as $\dd V^2$ while the 1-halo scales as $\dd V$ ; correspondingly the bispectrum has terms in $\dd V^3$, $\dd V^2$ and $\dd V$, thus high order statistics allow to better constrain the comoving volume by breaking degeneracies.


\section{Discussion}\label{Sect:discu}

The galaxy angular power spectrum thus contain valuable raw information on cosmological parameters in the highly non-linear regime dominated by the 1-halo term. The main finding of this article is that in this regime the information rises steeply with a slope comparable to that in the linear regime. This could in principle yield huge improvement to constraint on the Dark Energy equation of state.

One condition to realise these promises is to reach the ``small scale miracle'' which lies on scales $k>3$ Mpc$^{-1}$. This is currently out of reach of the best methods to predict the galaxy power spectrum such as the \texttt{matryoshka} emulator \citep{DonaldMcCann2021}, at least at the 1\% precision level. One can however hope to reach these scales in future as the needed improvement of reach is a factor<10 ; there is for instance a proposal that could reach these scales by \citep{Hannestad2019}.

Predicting matter statistics is however not a sufficient condition to realise the small scale miracle. To my knowledge, only galaxy and intensity mapping can measure this regime accurately, as for instance cosmic shear gets dominated by shape noise earlier on. So we further need a modeling of galaxies to these scales. The most promising prediction framework for this seems to be the halo model. There are issues often raised against the halo model but recent progresses could solve them. For instance the problem with mass conservation and large scales is solved in the Extended Halo Model (EHM) of \cite{Schmidt2016} ; the problem of imprecision in the transition regime between 2-halo and 1-halo can be solved by the Amended Halo Model (AHM) of \cite{Chen2019}. Furthermore, none of these issues affect the small scale miracle, which lies in the highly non-linear regime where the 1-halo term is dominant.

Another condition is to control the uncertainties associated to the impact of galaxy formation and baryonic physics. The impact of these uncertainties is partially studied in sections \ref{Sect:marginalisation} and \ref{Sect:caveats} and Appendix \ref{Sect:ext-astro}. A full assessment is however of a greater scope beyond this article. There is arguably uncertainty on whether we can control these uncertainties. These studies are thus left for future works.

On a positive note, although using the 1-halo term for cosmology is not (yet) usual in galaxy surveys, this has been done routinely in other surveys.
For example in the case of analyses of the thermal Sunyaev-Zel'dovich (tSZ) effect, the angular power spectrum is entirely dominated by the 1-halo term on all scales that have been observed to date. And cosmological constraints have been extracted out of the tSZ power spectrum for instance by \cite{Planck2013-SZmap,Planck2015-SZmap}. It has been shown that there is a high interest in analysing higher order statistics : for instance in \cite{Planck2013-SZmap} the tSZ bispectrum gave cosmological constraints of comparable power to those of the power spectrum, both analyses being in fact limited by uncertainties in the astrophysics of the cluster gas. Furthermore, \cite{Hurier2017} showed that the tSZ bispectrum, power spectrum and cluster counts have a great synergy that breaks degeneracies between cosmological and astrophysical parameters.

Furthermore, in the case of galaxy surveys \cite{Lacasa2019b} studied the galaxy angular power spectrum and showed that the inclusion of non-Gaussian covariance terms in the covariance decreases the degeneracies between cosmological and HOD parameters. \cite{Lacasa2016} also showed that the inclusion of cluster counts allows to break some of these degeneracies. So disentangling these two informations could be possible in the 1-halo dominated regime.

In conclusion there is high interest in pushing the analyses of galaxy clustering to scales of a few Mpc$^{-1}$ to 10 Mpc$^{-1}$, dominated by the 1-halo term. Especially considering that this regime will be measured "for free" with future high resolution surveys of the large scale structure.


\section*{Acknowledgements}
\vspace{0.2cm}

Part of this work was supported by funds of the D\'epartement de Physique Th\'eorique, Universit\'e de Gen\`eve. Part of this work was supported by a postdoctoral grant from Centre National d’Études Spatiales (CNES).
Some of the computations made use of the CLASS code \citep{Blas2011}.

\bibliographystyle{aa}
\bibliography{bibliography}

\appendix

\section{Extended astrophysical model and impact on the cosmological information content}\label{Sect:ext-astro}
This appendix studies a +4-parameter extension of the astrophysical base model, in order to gauge whether it impacts the extractable cosmological information after marginalisation.

\subsection{Extended model: presentation}\label{Sect:ext-mod-pres}
I first introduce two new parameters to give more freedom to the Halo Occupation Distribution. The first parameter is called $\alpha_1$ and it allows for redshift dependence of the high mass slope of the HOD:
\ba
\mathbb{E}\left[N_\mr{sat}|N_\mr{cen}=1\right]= \left(\frac{M}{M_\mr{sat}}\right)^{\alpha_\mr{sat}+\alpha_1 (1+z)} .
\ea
The fiducial value is $\alpha_1=0$, i.e. backward compatibility with the base model.\\
The second parameter is called $\beta_\mr{ratio}$ and it allows for redshift dependence of the central-satellite transition of the HOD:
\be
M_\mr{sat} = M_\mr{min} \times M_\mr{ratio} \, (1+z)^{\beta_\mr{ratio}}
\ee
The fiducial value is $\beta_\mr{ratio}=0$, i.e. compatible with the base model.\\
That makes a total of 9 HOD parameters in this extended model, hence it allows nearly as much freedom in the galaxy bias as leaving it as a free parameter in each of the 10 redshift bins. The HOD parameters also allow more freedom in the 1-halo term of the power spectrum.

The next parameter relates to the stochastic distribution of the number of galaxies in a halo. In the base model used in the main text, this was assumed to follow a Poisson distribution, as supported by simulations \citep{Kravtsov2004} and observations \citep{Yang2008}. In this extended model we allow the moments of this distribution to deviate from the Poisson prediction by a deviation parameter $\mathcal{A}_P$ \citep{Cacciato2013}:
\be
\lbra N_\mr{sat} (N_\mr{sat}-1) \rbra = \lbra N_\mr{sat} \rbra^2 \times  \mathcal{A}_P
\ee
so that
\be
\lbra N_\mr{gal} (N_\mr{gal}-1) \rbra = 2 \lbra N_\mr{sat} \rbra + \lbra N_\mr{sat} \rbra^2 \times  \mathcal{A}_P
\ee
The fiducial value is $\mathcal{A}_P=1$, again compatible with the base model. Following \cite{Cacciato2013}, a Gaussian 10\% prior can be imposed on this parameter.\\
$\mathcal{A}_P$ exclusively impacts the 1-halo term of the power spectrum, thus degrading constraints from it. Effectively it gives more freedom in how $C_\ell^{1h}$ probes the halo mass function.

The last parameter relates to the (average) spatial distribution of galaxies inside the halo. We follow the general idea of \cite{More2015} to allow this distribution to deviate from the dark matter distribution, but instead of doing it only for a fraction of central galaxies we allow it for all galaxies. This is both simpler and more conservative. The parameter is called $R_S$ and rescales the NFW profile:
\be
u_\mr{gal}(k|M,z) = u_\mr{NFW}(k\times R_S|M,z)
\ee
The fiducial value is $R_S=1$ (for compatibility with the base model). We will investigate the impact of a Gaussian 20\% prior on this parameter.\\
$R_S$ affects all halo model integrals of the form Eq.~\ref{Eq:unified-HM-integrals}, allowing more freedom in their scale dependence. In principle, this affects both the 2-halo and 1-halo term of the power spectrum. Indeed the 1-halo term is directly such an integral, and the 2-halo term contains the galaxy bias which is such an integral. In real life however, $R_S$ will only affect (=degrade) the constraining power from the 1-halo term. Indeed the introduced freedom will be relevant only on halo scales, where the 1-halo term has overtaken the 2-halo term.

In a way, the $\mathcal{A}_P$ and $R_S$ parameters can be argued to mimick the impact of baryonic physics, because they impact the amplitude and scale dependence of the 1-halo term by changing how many (pairs of) galaxies are present in a halo and where they distribute.

In summary, the extended model is fairly conservative, as it allows more freedom in the galaxy bias, and more importantly much more freedom in the amplitude and scale dependence of the 1-halo term.

\subsection{Impact on cosmological errors}

\begin{figure}[!ht]
	\begin{center}
		\includegraphics[width=\linewidth]{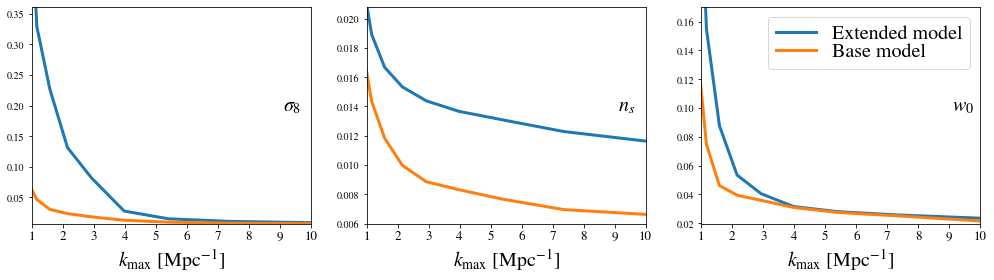}
		\caption{Marginalised error bars on $\sigma_8$, $n_S$ and $w_0$ as a function of $\kmax$, depending on the astrophysical model. Blue curve: with the extended model, without any prior on the additional parameter. Orange curve: with the base model of the main text.}
		\label{Fig:marginalised-errors-kmax-basevsext}
	\end{center}
\end{figure}

Figure \ref{Fig:marginalised-errors-kmax-basevsext} shows the marginalised error bars on $\sigma_8$, $n_S$ and $w_0$ as in Fig.~\ref{Fig:marginalised-errors-kmax} but comparing the base model and the extended model. Note that here, no prior is imposed at all on any of the additional parameter, so this is an \"{u}ber-conservative case.

\begin{figure}[!ht]
	\begin{center}
		\includegraphics[width=\linewidth]{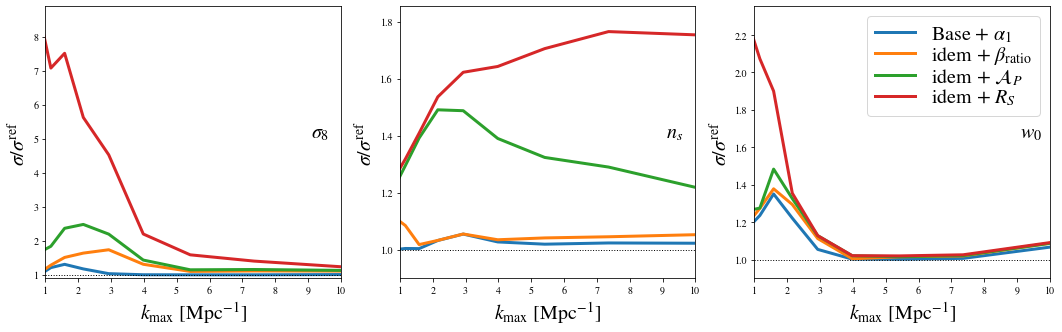}
		\caption{Ratio of the marginalised error bars from a given model to those of the base model (ref). Considered models: \textit{(blue)} Base model + $\alpha_1$, \textit{(orange)} Base model + $\alpha_1$ + $\beta_\mr{ratio}$, \textit{(green)} Base model + $\alpha_1$ + $\beta_\mr{ratio}$ + $\mathcal{A}_P$, and \textit{(red)} Base model + $\alpha_1$ + $\beta_\mr{ratio}$ + $\mathcal{A}_P$ + $R_S$ = extended model.}
		\label{Fig:marginalised-errors-ratios}
	\end{center}
\end{figure}

Generally, the constraints take a large hit at low $k_\mr{max}$ (I checked this is even more the case on larger scales: $k=0.1-1$ Mpc$^{-1}$), especially for $\sigma_8$ which can degrade by up to a factor 8. For $n_S$, the  impact persists to high $k_\mr{max}$, though we still get $\sim$1\% constraints, which is fairly decent. For $\sigma_8$ and $w_0$ however, the impact seems to decrease largely as we increase $k_\mr{max}$.\\
This last point is not the most clearly visible, as at high $k_\mr{max}$ the curves are close but also have small values. Furthermore we would like to see more details on which parameter(s) drive this increase of error bars. This is why Fig.~\ref{Fig:marginalised-errors-ratios} shows the ratio of the errors to those of the base model, for different extensions.

With four parameters to set on/off, there should be $2^4$ curves to be exhaustive. But for simplicity and to avoid overcrowding the plot, only four models were chosen: extending the base model by one parameter at a time, in the order they are presented in Sect.~\ref{Sect:ext-mod-pres}. One sees that $R_S$ is generally the parameter which degrades the most the error bars. Exceptions are: $n_S$ at $k_\mr{max}< 2$ Mpc$^{-1}$ where $\mathcal{A}_P$ has the most significant impact, and $w_0$ at $k_\mr{max} > 2$ Mpc$^{-1}$ where $\alpha_1$ has the most significant impact. The figure also confirms the high $k_\mr{max}$ behaviour hinted in Fig.~\ref{Fig:marginalised-errors-kmax}: when pushing to small scales, the impact of the additional parameters becomes less relevant for $\sigma_8$ and $w_0$. For example at  $k_\mr{max} = 10$ Mpc$^{-1}$, the increase of error bar is only of 24\% for $\sigma_8$ and 9\% for $w_0$. For $n_S$ the impact is more consequent and does not decrease with $k_\mr{max}$ ; it reaches 75\% at  $k_\mr{max} = 10$ Mpc$^{-1}$. 

Still, the resulting constraints are comparable to the original ones and are fairly decent. One may hope to further improve them by setting priors on the additional parameters.

\subsection{Effect of realistic priors}

The previous sub-section was \"{u}ber-conservative in not setting any prior on the additional parameter. However we can realistically set a Gaussian prior of width 10\% on $\mathcal{A}_P$ and a Gaussian prior of width 20\% on $R_S$. Furthermore, the whole article assumed $H_0$ to be a free parameter, which degrades the cosmological constraints since galaxy clustering alone has extremely poor constraints on $H_0$. In reality, one would take a prior on $H_0$ from CMB or supernovae data\footnote{Given the current Hubble tension, which central value to take is an issue for actual analysis. However here this is not an issue since Fisher analysis is only affected by the width of the prior (assuming the fiducial model is correct of course).}. So I also consider the case for a prior on h, of width given by the Planck 2018 constraints \citep{Planck2018-cosmo}.

\begin{figure}[!ht]
	\begin{center}
		\includegraphics[width=\linewidth]{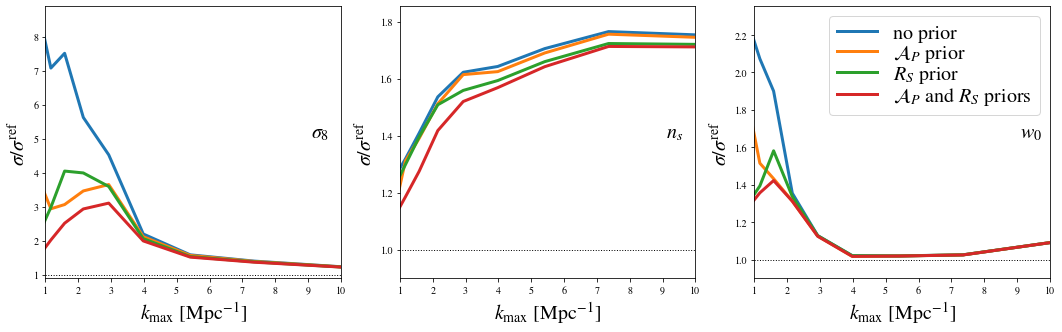}
		\caption{Ratio of the marginalised error bars to those of the base model (ref), when imposing different realistic priors.}
		\label{Fig:marginalised-errors-ratios-priors}
	\end{center}
\end{figure}

Let us first focus on the astrophysical parameters. Figure~\ref{Fig:marginalised-errors-ratios-priors} shows the impact of the priors on $\mathcal{A}_P$ and $R_S$ on the cosmological errors compared to the base model.

These realistic priors provide a spectacular improvement of the error on $\sigma_8$ at low $k_\mr{max}$, dividing the error by up to a factor 3.5 at $k_\mr{max} \sim 1$ Mpc$^{-1}$. As $k_\mr{max}$ increases, the impact of the priors lessens, to become basically insignificant at $k_\mr{max} > 4$ Mpc$^{-1}$. This is however a regime where the errors gradually converge to that of the base model. The results are qualitatively similar for $w_0$, with the impact of priors saturating even more quickly and the error being already close to that of the base model. The results for $n_S$ are disappointing however: the impact of the priors is meh.

Second, I studied the impact of an $H_0$ prior. To avoid comparing apples and oranges, I defined the new reference to be the base model with $H_0$ prior. I found some impact of the $H_0$ prior on $\sigma_8$ and $w_0$: it somewhat improves the error bars at low $k_\mr{max}$, though the priors on $\mathcal{A}_P$ and $R_S$ have a much higher effect there. For $n_S$ the situation is more interesting: the $H_0$ prior improves the error ratio significantly and stops it from increasing with $k_\mr{max}$. For instance the error ratio extended/base (both with the $H_0$ prior) reaches $\sim$40\% at $k_\mr{max} = 10$ Mpc$^{-1}$.

In conclusion, for $\sigma_8$ and $w_0$ the priors on $\mathcal{A}_P$ and $R_S$ are the most needed ; they do help a lot where the extended model suffers the most in comparison with the base model (low $k_\mr{max}$). For $n_S$ however, it is the $H_0$ prior that is the most needed ; it does help to stabilise the extended/base ratio, though there is some room for improvement.

\subsection{Changing the prior size}
Here we ask whether having better priors can help to improve our constraints.

\subsubsection{General formula: impact of a Gaussian prior on Fisher constraints}\label{Sect:App-formula-Fisher-prior}
This section will make use of the Sherman-Morrison formula \citep{Sherman1950,Bartlett1951}, which gives the impact on matrix inversion of a rank 1 update:
\ba\label{Eq:Sherman-Morrison}
\left(A+U V^T\right)^{-1} = A^{-1} - \frac{A^{-1} \cdot U \cdot V^T \cdot A^{-1}}{1 + V^T \cdot A^{-1} \cdot U},
\ea
where $A$ is a square matrix, $U$ and $V$ are vectors, and $T$ means the transpose.

In Fisher forecasts, applying a Gaussian prior of width $\sigma^2$ on a model parameter with index $i_0$ is done by simply adding to the original Fisher matrix $F$ a second matrix $F^\mathrm{prior}$ such that
\be
F^\mathrm{prior}_{i,j} = \frac{1}{\sigma^2} \ \delta_{i,i_0} \ \delta_{i_0,j}
\ee
This matrix is of rank 1 and can be written as $F^\mathrm{prior} = U_\mathrm{p} U_\mathrm{p}^T$ with
\be
\left(U_\mathrm{p}\right)_{i} = \frac{1}{\sigma} \ \delta_{i,i_0}
\ee
Then noting the new Fisher matrix $\tilde{F}=F+F^\mathrm{prior}$, the Sherman-Morrison formula gives
\be
\tilde{F}^{-1} = F^{-1} - \frac{F^{-1} \cdot U_\mathrm{p} \cdot U_\mathrm{p}^T \cdot F^{-1}}{1+ U_\mathrm{p}^T \cdot F^{-1} \cdot U_\mathrm{p}}
\ee
which yields
\be\label{Eq:Fisher-mpact-prior}
\tilde{F}^{-1}_{i,j} = F^{-1}_{i,j} - \frac{F^{-1}_{i,i_0} F^{-1}_{i_0,j}}{\sigma^2+ F^{-1}_{i_0,i_0} }
\ee
or written another way
\be
\tilde{F}^{-1} = F^{-1} - \frac{S(i_0) \cdot S(i_0)^T}{\sigma^2+ F^{-1}_{i_0,i_0} }
\ee
where $S(i_0)$ is the column vector (slice) of $F^{-1}$ at position $i_0$: $S(i_0)_i = F^{-1}_{i,i_0}$.\\
The function smoothly interpolates between the model where the parameter with index $i_0$ is fixed at its fiducial value ($\sigma=0$) and the original model where it is entirely free ($\sigma \rightarrow \infty$). The difference between the two models is:
\be
\Delta \tilde{F}^{-1}_{i,j} = \frac{F^{-1}_{i,i_0} F^{-1}_{i_0,j}}{F^{-1}_{i_0,i_0} }
\ee
In terms of constraining power, the midpoint between the two models is reached for $\sigma=\sqrt{F^{-1}_{i_0,i_0}}$. Note that this midpoint position does not depend on the parameter $i$ considered. It happens when the prior has the same width as the data constraint.

One could further generalise the formula above to allow for simultaneous priors on several parameters (even correlated priors), by using the Woodbury matrix identity \citep{Woodbury1950}. Indeed this identity gives the impact on matrix inversion of an update with arbitrary rank. We refrain to explore these complications here and will only study one prior at a time in the next subsection.

\subsubsection{Application}
\begin{figure}[!ht]
	\begin{center}
		\includegraphics[width=\linewidth]{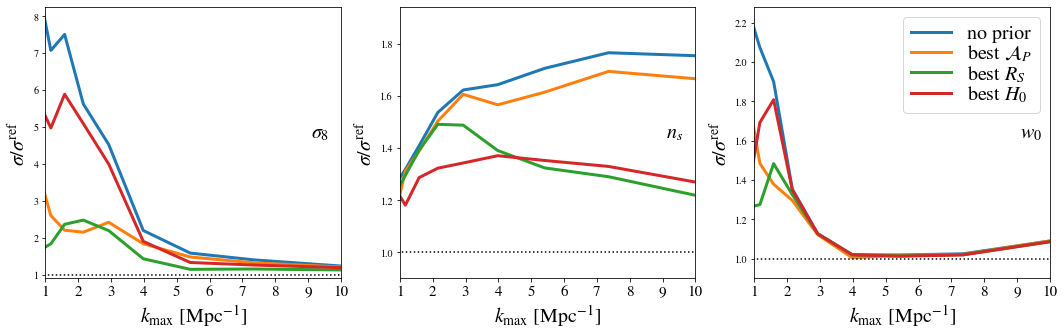}
		\caption{Ratio of the marginalised error bars to those of the base model (ref), when imposing perfect priors on different parameters.}
		\label{Fig:marginalised-errors-ratios-bestpriors}
	\end{center}
\end{figure}

A first question we need to ask is \textit{when} can prior help. This is partially answered by Fig.~\ref{Fig:marginalised-errors-ratios-priors}, but one may be worried that the figure does not show the full potential of priors. So Fig.~\ref{Fig:marginalised-errors-ratios-bestpriors} shows the same error bar ratios but now imposing the best possible priors (i.e. $\sigma=0$ in Eq.~\ref{Eq:Fisher-mpact-prior}) on $\mathcal{A}_P$ and $R_S$. The impact of prior on $H_0$ is also shown for curiosity. 

As in Fig.~\ref{Fig:marginalised-errors-ratios-priors}, we see that for $\sigma_8$ and $w_0$, these priors are most interesting at low $k_\mr{max}$ ; on smaller scales they become unimportant. For $n_S$ however, we see that the priors are the most interesting at high $k_\mr{max}$, in particular for $R_S$ (and $H_0$).

In these conditions, the ultimate question is: what prior size would we need to feel these improvements ?\\
The main tool to answer this question is the remark at the end of Sect.~\ref{Sect:App-formula-Fisher-prior}: the midpoint is reached when $\sigma=\sqrt{F^{-1}_{i_0,i_0}}$, i.e. when the prior size is equal to the data constraint.\\
For $H_0$ the question is then readily answered: the current precision is already well enough on all scales. Indeed, we have currently 1\% precision on $H_0$, and the constraint from galaxy clustering is ridiculously worse than that (500\% at best). So we are in the regime $\sigma \ll \sqrt{F^{-1}_{i_0,i_0}}$, the constraint are basically saturated to the $\sigma=0$ value (=fixing $H_0$).\\
For $\mathcal{A}_P$ and $R_S$ the answer will depend on the cosmological parameter considered and on scale. First, for $\sigma_8$ and $w_0$, we noted previously that the interesting regime is at low $k_\mr{max}$. For instance at $k_\mr{max}=1.2$ Mpc$^{-1}$ the data constraint on $\mathcal{A_P}$ is 47\% and on $R_S$ is 16\%. So we are in a regime $\sigma < \sqrt{F^{-1}_{i_0,i_0}}$: the current prior are good. Improving these priors would improve a bit cosmological errors, but if it were not possible this would not be a show-stopper. Second, for $n_S$, we noted previously that the interesting regime is at high $k_\mr{max}$. For instance at $k_\mr{max}=10$ Mpc$^{-1}$ the data constraint on $\mathcal{A_P}$ is 3.3\% and on $R_S$ is 5.6\%. We saw on Fig.~\ref{Fig:marginalised-errors-ratios-bestpriors} that prior on $\mathcal{A_P}$ is uninteresting for $n_S$, so let us concentrate on $R_S$. Here we are in a regime where the current prior has $\sigma > \sqrt{F^{-1}_{i_0,i_0}}$, i.e. it is not good enough, it fails to improve $n_S$ (as can also be seen by comparing Fig.~\ref{Fig:marginalised-errors-ratios-priors} and Fig.~\ref{Fig:marginalised-errors-ratios-bestpriors}). To mitigate the impact of the model extension, we should aim for having external knowledge to constrain $R_S$ at the $< 5$ \% level.

\begin{figure}[!ht]
	\begin{center}
		\includegraphics[width=\linewidth]{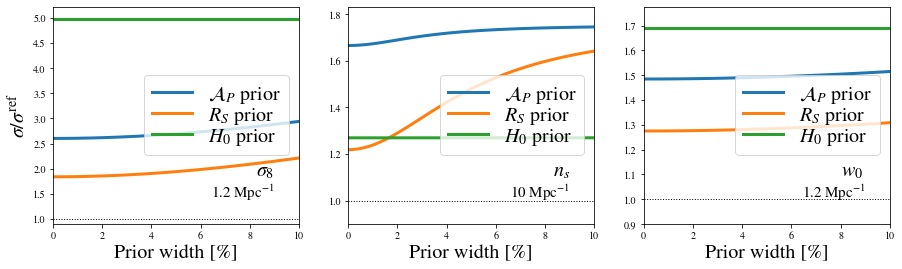}
		\caption{Ratio of the marginalised error bars to those of the base model (ref), depending on the prior width. $k_\mr{max}$ is fixed for each parameter to a regime where the priors matter: $\sigma_8$ and $w_0$ are at a low $k_\mr{max}$, while $n_S$ is at a high $k_\mr{max}$. Current knowledge on $H_0$ is already largely enough to saturate the prior impact, hence the $H_0$ curves are flat. For $\mathcal{A}_P$ and $R_S$, the current knowledge is informative enough for $\sigma_8$ and $w_0$, but better precision at the <5\% level would be nice for futuristic constraints on $n_S$.}
		\label{Fig:marginalised-errors-ratios-priorwidth}
	\end{center}
\end{figure}

In conclusion, the outlook is optimistic to me. For near-future analyses staying at low $k_\mr{max}$, the current priors on $\mathcal{A}_P$ and $R_S$ are nearly enough to extract all the constraining power on $\sigma_8$ and $w_0$, and they are not needed for $n_S$. For more futuristic analyses at high $k_\mr{max}$, priors on $\mathcal{A}_P$ and $R_S$ will not be needed for $\sigma_8$ and $w_0$. A <5\% $R_S$ prior will be needed to unlock the full constraining power on $n_S$, but (i) this is not a show-stopper, the degradation is at worst 50\% (no $R_S$ prior at all, still with a $H_0$ prior), and (ii) by the time, it is conceivable the community will have improved on current knowledge and reached this precision.

\end{document}